\documentclass[%
twocolumn,
 amsmath,amssymb,
aps,
pra,
]{revtex4-1}

\usepackage{graphicx,epsfig,epstopdf}
\usepackage{amsmath}
\usepackage{color}
\usepackage{caption}
\usepackage{subcaption}
\usepackage{braket}

\def\e{\begin{equation}}
\def\f{\end{equation}}
\def\_#1{{\bf #1}}
\def\.{\cdot}

\begin{document}

\title{Omnidirectional perfect acoustic cloak realized by homogeneous materials  } 

\author{Mohammad Hosein Fakheri}
\author{Hooman Barati Sedeh}
\author{Ali Abdolali}
\email{Abdolali@iust.ac.ir}
\affiliation{
	Applied Electromagnetic Laboratory, School of Electrical Engineering, Iran University of Science and Technology, Tehran, 1684613114, Iran}
\begin{abstract}
Acoustic cloaks derived by coordinate transformation have opened up a new field of considerable interest in the last two decades. However, since perfect omnidirectional acoustic cloak relies on inhomogeneous and anisotropic materials that posses extreme values in certain regions, this cloak was deemed impossible to be attained even with metamaterials. Recently, our group was competent to introduce a new kind of extreme acoustic material, named as acoustic null medium (ANM), for attaining various devices. In this letter, an extreme mapping function is exploited to achieve an omnidirectional acoustic cloak. The attained extreme homogeneous material through this transformation becomes ANM that can satisfy the extreme parameters requirement and as a result, can hide the
object from any incident acoustic waves. Several numerical simulations were carried out to demonstrate the capability of the propounded cloak in cloaking objects from sound fields. The propounded work will pave the way towards cloaking arbitrary objects from any incident angle and obviate the conventional challenges regarding this device.
\end{abstract}

\maketitle

\section{introduction}
The concept of transformation optics (TO), paves the way toward  
controlling electromagnetic (EM) waves in an unprecedented manner and results in many novel devices, which were deemed impossible such as invisibility
cloak \cite{pendry2006controlling,fakheri2017carpet}. Soon after the introduction of TO, this approach was extended to other physical systems, including acoustics and named as transformation acoustics (TA) \cite{cummer2007one}. Introducing TA as a mean to manipulate acoustic waves soon gave rise to acoustic cloak, which is a coating layer that surround the object and makes it invisible from sound fields \cite{chen2010acoustic}. However, beside the inhomogenity and anisotropy issue, the main drawback of a perfect cloak is its demand for extreme material, which restrict its applicability to theoretical investigations. This need is a direct result of mapping a point from volumeless dimension (0D) to a two-dimensional (2D) space, which is the fundamental transformation of a perfect cloak \cite{cummer2007one}. To obviate this problem, several procedures have been proposed including ground plane cloak (also known as carpet cloak), which is capable of restoring the acoustic scattering signature of the object in a manner that the incident acoustic wave is scattered from a plain surface \cite{li2008hiding,ma2009compact,kallos2009ground,ma2010three}. Since the mentioned cloak is designed via establishing a transformation function, which links a two-dimensional (2D) flat line segment to a 2D curved line, it can obviate the demand of extreme materials parameters. Not long after the introduction of ground plane cloaks, unidirectional acoustic cloaks, which are devices that are designed for exhibiting cloaking functionality only for a specific incident direction, were propounded based on TA methodology, that do not need extreme materials or limited to function only on  plain surfaces \cite{xi2009one,landy2013full}.  Although the mentioned cloaks were firstly proposed by  performing quasi-conformal mapping function which can  reduce the anisotropy of the obtained materials, the inhomogenity of the materials will lead to a difficult fabrication process.  In addition, neglecting the weak anisotropy will give rise to lateral shift in the reflected wave, which will degrade the efficiency of the designed cloaks \cite{zhang2010lateral}. Furthermore, the mentioned cloak can function under a specific incident direction. That is, if the incident angle is changed, which is a realistic assumption, the designed cloaks will malfunction and will scatter the incident waves \cite{landy2013full}. \par  
In this letter, we show that omnidirectional acoustic cloaks attained by applying linear transformation function are feasible in practice, although they will posses extreme materials. Two different materials are obtained from this transformation, which one of them is a simple anisotropic material and the extreme one is acoustic null medium (ANM) that is recently proposed by our group and Fei \textit{et al.} group \cite{fakheri2018arbitrary,sun2019full,li2017acoustic}. Several simulations were performed and their results demonstrated a high level transmission of acoustic wave and total scattering reduction,  which corroborate the effectiveness of the proposed approach. 
\section{Theoretical formulation}
The schematic of the acoustic omnidirectional cloak is demonstrated in Fig.1 (a). It is notable to mention that the propounded idea could be easily extended to three dimensional (3D) case with any polygonal geometry; however, here we will present its 2D results with square shape geometry as a demonstration of our idea.  In contrary to the previous reported cloaks which a point in the virtual space is mapped into a circle in the physical space, here the cloak region has divided into different regions and for each domain an affine transformation, is used as shown in Fig.1 (b). 
\begin{figure}
	\centering
	\includegraphics[width=0.5\textwidth]{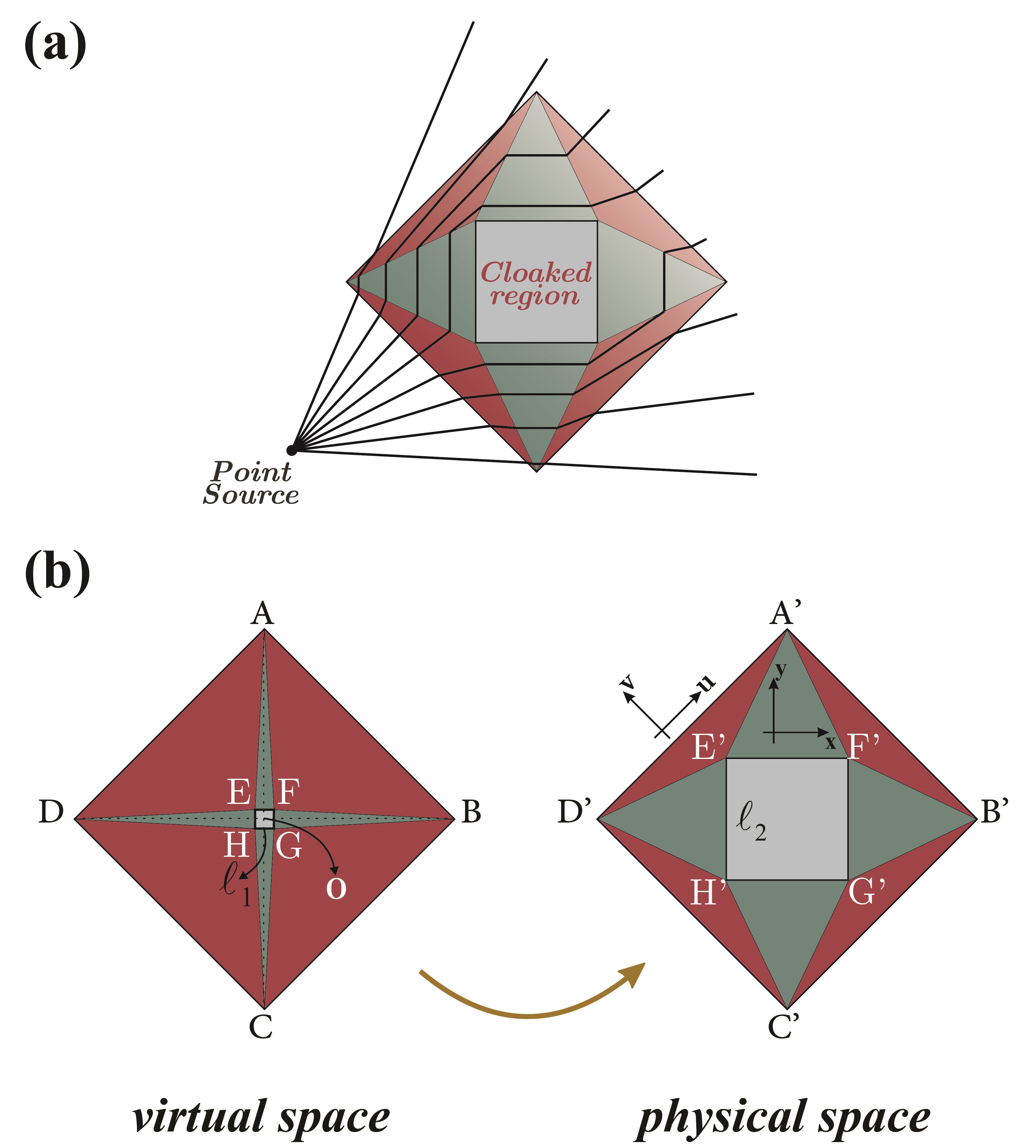}
	\caption{(a) The Schematic of the acoustic omnidirectional cloak obtained by exploiting affine transformation. (b) The utilized transformation function between virtual and physical spaces.The black lines indicates the flow of acoustic waves before and after entering the cloaked region.}
	\label{fgr:fig1}
\end{figure} 
The space between the square with side length of $L$ (i.e., \textit{ $ A B C D$}) in the virtual space is transformed to the same square with the same length in physical space(i.e., \textit{$A^\prime B^\prime C^\prime D^\prime$}), while at the same time inner square with the side of \textit{$l_1$} (i.e., \textit{$E F G H$})in the virtual space is mapped to a larger square with the side of \textit{$l_2$} (i.e., \textit{$E^\prime F^\prime G^\prime H^\prime$}) in the physical space. Without the loss of generality, one can assume that the inner square side length is approaching to zero (i.e., $l_1 \rightarrow 0$ ). Therefore, under this assumption, the triangles $\triangle OAB$, $\triangle OBC$, $\triangle OCD$ and $ODA$ in the virtual
space (Fig. 1b) are transformed to triangles $\triangle F^ \prime A^\prime B^\prime$, $\triangle G^\prime B^\prime C^\prime $, $\triangle H^\prime C^\prime D^\prime$ and
$\triangle E^\prime D^\prime A^\prime $ in the physical space (Fig. 1b), respectively. Meanwhile, the lines $AO$,
$BO$, $CO$ and $DO$ in the virtual space must also be transformed to
triangles $\triangle A^\prime E^\prime F^\prime $, $\triangle B^\prime F^\prime G^\prime $, $\triangle C^\prime G^\prime H^\prime $ and $\triangle D^\prime H^\prime E^\prime$ in the physical space, respectively. On the other hand, the mapping between two coordinate systems (i.e. virtual
space $(x, y,z)$ and physical space $(x^\prime, y^\prime,z^\prime)$) can be characterized in
terms of the Jacobian matrix $\Lambda=\partial(x^\prime,y^\prime,z^\prime)/(x,y,z)$ \cite{cummer2007one}. Then, the mass density tensor and bulk modulus of the transformed media are
given by 
\begin{align}
&\bar{\bar{\rho}}^\prime= det(\Lambda) (\Lambda^{-1})^T \rho_0 (\Lambda)^{-1}\\ \nonumber
&\kappa^\prime=det(\Lambda) \kappa_0
\end{align}
whence $\rho_0$ and $\kappa_0$ are the mass density and bulk modulus of the background medium which is assumed to be air. Thus, following the above-mentioned points, the necessitating materials for each of the green triangles will be achieved as (the details are provided in the supplementary file)
\begin{subequations}
	\begin{equation}
	\label{eq-a}
	\frac{\rho^\prime_{L,R}}{\rho_0}=
	\begin{bmatrix}
	\infty&
	0&
	0 \\
	0 &
	
	0  &
	0\\
	0 &
	0 &
	\infty  
	\end{bmatrix}, \kappa^\prime_{L,R}/\kappa_0=\infty 
	\end{equation}
	\begin{equation}
	\frac{\rho^\prime_{T,B}}{\rho_0}=
	\begin{bmatrix}
	0&
	0&
	0 \\
	0 &
	
	\infty  &
	0\\
	0 &
	0 &
	\infty  
	\end{bmatrix}, \kappa^\prime_{T,B}/\kappa_0=\infty 
	\end{equation}
\end{subequations}
where $L,R$ stands for left and right green triangles (i.e.,  $\triangle D^\prime E^\prime H^\prime$ and $\triangle F^\prime B^\prime G^\prime$ ), while $T,B$ indicates the top and bottom ones (i.e., $\triangle A^\prime E^\prime F^\prime$ and $\triangle H^\prime G^\prime C^\prime$).
Following the same procedure, the necessitating materials will be achieved in their local coordinate system $(u,v,z)$ as
\begin{equation}
\frac{\rho^\prime}{\rho_0}=
\begin{bmatrix}
\chi&
0&
0 \\
0 &

\frac{1}{\chi} &
0\\
0 &
0 &
\chi
\end{bmatrix}, \kappa^\prime/\kappa_0=\chi
\end{equation}
where $\chi=1- l_2/[L\sin (\pi/4)]$.
Compared with the previous cloak prototypes with inhomogeneous and anisotropic extreme material parameters, the square cloak is simplified to only two homogeneous material . The latter (i.e., Eq. (3)) is a simple diagonal anisotropic mass density tensor which could be easily implemented via acoustic metamaterials \cite{chen2007acoustic,li2004double}. While, the former (i.e., Eq. (2)) is a new material which has been recently proposed by our group and Fei \textit{et al.} \cite{fakheri2018arbitrary,li2017acoustic} and named as acoustic null medium (ANM) to achieve different functionalities. In fact, ANM is a material with extreme
anisotropic parameters, which is obtained by mapping a volumeless
line (or surface in 3D case)in the virtual space to a surface (or a volume in 3D scenario) in the physical space. This will make the incident wave to be mapped from one interface to another point-to-point, without any distoration or phase accumulation. \par 
\section{Numerical simulations}
To demonstrate the capability of the proposed cloak, several numerical simulations were carried out via COMSOL Multiphysics. For all the propounded simulations, the operative frequency is $f=10 kHz$ and the dimensions of the structure are set to be $L=5.66\lambda$ and $l_2=1.88\lambda$, respectively. The functionality of the cloak will be examined under two different acoustic illumination of plane wave and point source radiations. For each of the presented case the angle of incidence is varied to show the independency and omnidirectional behavior of the designed cloak. The first example is dedicated to the case of plane wave illumination with different incident angles of $\theta_{inc}=[0^\circ, 30^\circ, 60^\circ, 90^\circ ]$ as shown in Fig.2.

\begin{figure}[t]
	\centering
	\includegraphics [width=0.5\textwidth]{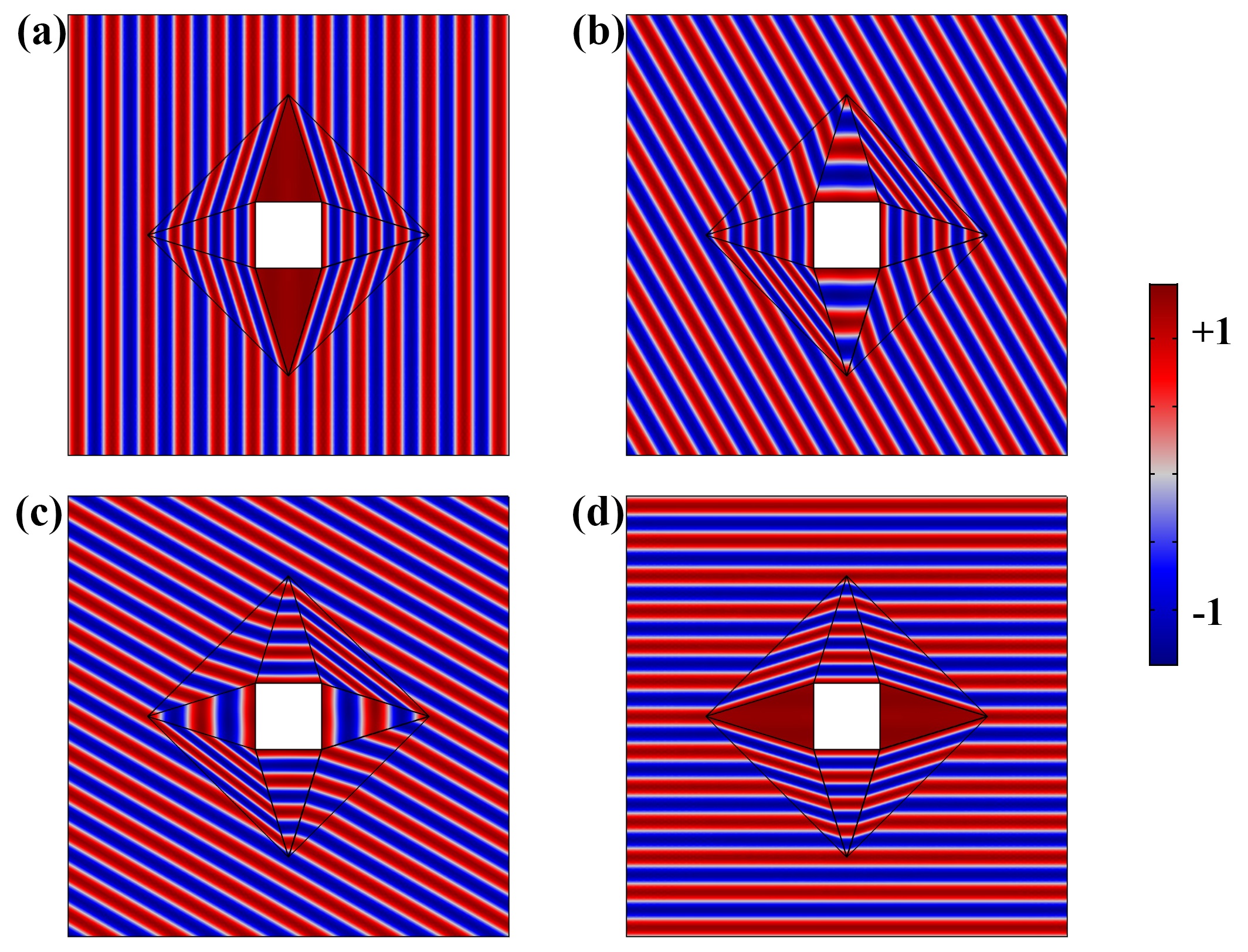}
	\caption{ The pressure distribution of the designed acoustic cloak under different incident angles of a) $\theta=0^\circ$, b) $\theta= 30^\circ$ , c) $\theta= 60^\circ$, d) $\theta= 90^\circ$}
	\label{fgr:fig2}
\end{figure}
As it is evident from this figure, the
forward and backward acoustics scatterings are totally suppressed
by covering the object with the designed cloak, while at the same time both of the phase and amplitude of the acoustic wave are well reconstructed. In addition, the conventional problem of TA-based cloak, which was the limited angle of view, is also obviated since the extreme material which obtained from this approach is ANM and could be easily implemented \cite{li2017acoustic,fakheri2018arbitrary}. In addition, in realistic situations, the illumination acoustic source might not be a plane wave but instead a point source. As it is shown in Fig.3 (a), when an object is being impinged by an acoustic point source, the acoustic fields will be perturbed both in backward and forward directions.

However, when the designed acoustic square shape cloak is utilized, the acoustic waves smoothly pass around the object and be reconstructed after leaving the cloak region with out showing any discrepancy in the near field distribution as shown in Fig.3 (b). Therefore, according to the above-mentioned discussions and results, it could be understand that the propounded square shape cloak is competent to perfectly hide the objects from acoustic waves omnidirectionally. 
\begin{figure}[h]
	\centering
	\includegraphics[width=0.5\textwidth]{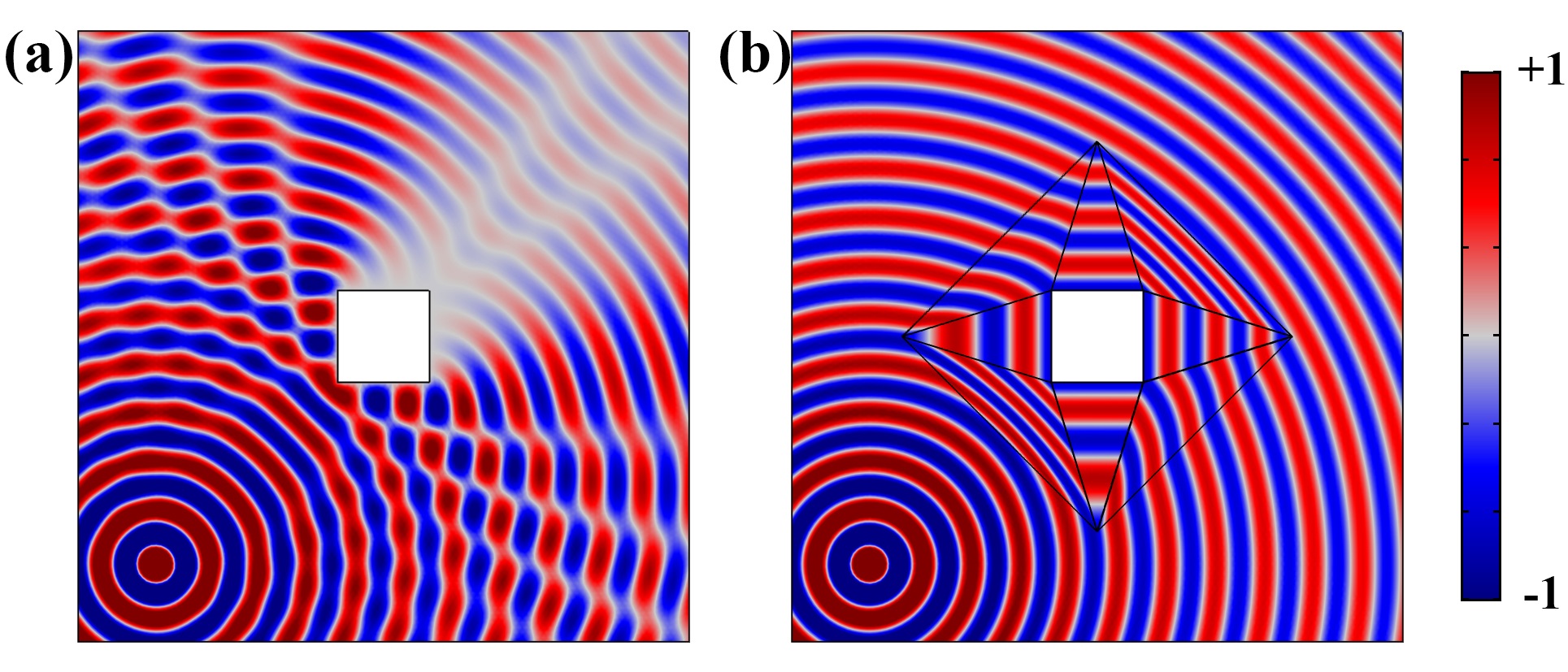}
	\caption{ The pressure distributions of an object under the illumination of an acoustic point source. a) When the designed cloak is not utilized. b) When the square shape cloak is exploited.}
	\label{fgr:fig3}
\end{figure}
\section{Conclusion}
In conclusion, we design and realize the first omnidirectional acoustic cloak based on extreme homogeneous material that is named as ANM, which is capable of hiding objects with different shapes in a square shape domain. Both of the numerical and
realization results clearly demonstrate the good performance
of omnidirectional invisibility of the present cloak, verifying
the correctness of our design. The present work provides
an important guidance toward the realization of complex
functionalities with acoustic metamaterials, and will motivate various
transformation acoustic based devices for practical applications, such as concentrators, rotators, and acoustical
illusion apparatuses.

\bibliography{Ref}

\end{document}